\documentclass[12pt,preprint]{aastex}
\usepackage{emulateapj5}
\usepackage{epsfig}

\newcommand{\etal}{\mbox{et al.}}

\newcommand{\degree}{$^\circ$}

\newcommand{\rxte}{{\it RXTE}}

\newcommand{\ksxrb}{\mbox{KS 1731$-$260}}
\newcommand{\aqlxone}{\mbox{Aql X-1}}
\newcommand{\mxbgc}{\mbox{MXB 1743$-$29}}
\newcommand{\sixb}{\mbox{4U 1636$-$536}}
\newcommand{\sevenb}{\mbox{4U 1702$-$429}}
\newcommand{\slowb}{\mbox{4U 1728$-$34}}
\newcommand{\novab}{\mbox{4U 1608$-$522}}
\newcommand{\mxbecl}{\mbox{MXB 1659$-$298}}
\newcommand{\dipb}{\mbox{4U 1916$-$053}}

\newcommand{\saxmsp}{\mbox{SAX J1808.4$-$3658}}



\shortauthors{Muno \etal}
\shorttitle{Neutron Star Rotation and X-ray Bursts}

\begin{document}
\title{The Effect of Neutron Star Rotation on the Properties of 
Thermonuclear X-ray Bursts}
\author{Michael P. Muno,\altaffilmark{1,2}  
Duncan K. Galloway, \& Deepto Chakrabarty\altaffilmark{3}}
\affil{Center for Space Research,
Massachusetts Institute of Technology, Cambridge, MA 02139}
\email{muno,deepto,duncan@space.mit.edu}
\altaffiltext{1}{Current Address: Department of Physics and Astronomy, University of California, Los Angeles, CA 90095}
\altaffiltext{2}{Hubble Fellow}
\altaffiltext{3}{ALSO Department of Physics, Massachusetts Institute 
of Technology, Cambridge, MA 02139}

\begin{abstract}
Using observations with the {\it Rossi X-ray Timing Explorer}, we  
previously showed that millisecond oscillations occur preferentially in 
thermonuclear X-ray bursts with photospheric 
radius expansion from sources rotating near 600~Hz, while they occur with 
equal likelihood in X-ray bursts with and without radius expansion for 
sources rotating near 300~Hz. In this paper, we use a larger sample of 
data to demonstrate that the 
detectability of the oscillations is not directly determined by the properties 
of the X-ray bursts. 
Instead, we find that (1) the oscillations are observed almost exclusively 
when the accretion rate onto the neutron star is high, but that (2) 
radius expansion is only 
observed at high accretion rates from the $\simeq 600$~Hz sources, whereas it
occurs only at low accretion rates in the $\simeq 300$~Hz sources. The 
persistent millisecond
pulsars provide the only apparent exceptions to these trends.
The first result might be explained if the oscillation amplitudes are 
attenuated at low accretion rates by an extended electron corona. 
The second result indicates that the rotation period of the neutron star 
determines how the burst properties vary with accretion rate, possibly through
the differences in the effective surface gravity or the strength of the
Coriolis force. 
\end{abstract}


\section{Introduction}

Nearly-coherent millisecond brightness oscillations with frequencies between 
270--620~Hz
have been observed during thermonuclear X-ray bursts from 13 neutron star 
low-mass X-ray 
binaries (see Strohmayer \& Bildsten 2003, for a review). Two of these systems
also show persistent millisecond pulsations at the same frequencies 
in their non-burst emission
\citep{cha03,str03}. It is therefore
 widely accepted that the oscillations result from brightness patterns
that form on the surfaces of these rapidly-rotating neutron stars during 
X-ray bursts, thus probing two very 
different pieces of physics: the distribution of neutron star spin 
frequencies \citep[e.g.,][]{wz97,bil98a,cha03}, and how unstable nuclear 
burning 
proceeds on a neutron star's surface (e.g., Nath, Strohmayer, \& Swank 2002; 
Spitkovsky, Levin, \& Ushormirsky 2002; Muno, \"{O}zel,
\& Chakrabarty 2003b).

In the sources that exhibit the millisecond burst oscillations, 
they are only detected from about half of all X-ray bursts. 
We previously showed that if a distinction is
drawn between sources rotating at $\simeq300$~Hz and $\simeq600$~Hz based on 
the timing properties of their persistent emission 
\citep[see][for a review]{vdk00}, 
then there is a physical difference in the properties of the
bursts that exhibit oscillations \citep{mun01}. 
The fast, $\simeq$600~Hz oscillations almost always occur during the 
strongest X-ray bursts, during
which the photosphere of the neutron star is driven to a large radius 
by radiation pressure. In contrast,
the slow, $\simeq 300$~Hz oscillations occur with equal likelihood in bursts 
with and without
photospheric radius expansion. Observational selection 
effects or differences in viewing angles cannot by themselves produce the 
correlations between burst properties and the presence of oscillations. 

The proposed explanation for this difference has two parts 
\citep[][]{fra01,mun01}: (1) that 
burst oscillations are almost exclusively detected when the accretion 
rates onto the neutron stars are
relatively high ($\sim 0.1\dot{M}_{\rm Edd}$), even though the X-ray bursts 
themselves have also been observed at significantly lower accretion rates, 
and (2) that the properties of bursts at these high accretion rates are 
different in the fast and slow rotators. This explanation was motivated
largely by observations of only two systems (\slowb\ and \ksxrb), 
and still requires 
confirmation with a larger sample of sources. After seven years of 
operation, 
there is now sufficient data in the archive of observations taken with the 
{\it Rossi X-ray Timing Explorer} (\rxte) Proportional Counter 
Array \citep[PCA;][]{jah96} to examine this hypothesis for several sources. 
Therefore, in this paper we examine how the presence of millisecond
oscillations and of photospheric radius expansion in thermonuclear 
X-ray bursts are related to the persistent accretion 
rates onto the neutron stars.

\section{Observations and Results}

We obtained public \rxte\ PCA
observations of 12 out of 13 
sources of millisecond burst oscillations.\footnote{We omitted \mxbgc\ from 
our analysis because its persistent 
emission cannot be isolated from the several sources within 1\degree\ of the
Galactic center.}
We searched for X-ray bursts using the using the algorithm 
\begin{figure*}[thb]
\centerline{\epsfig{file=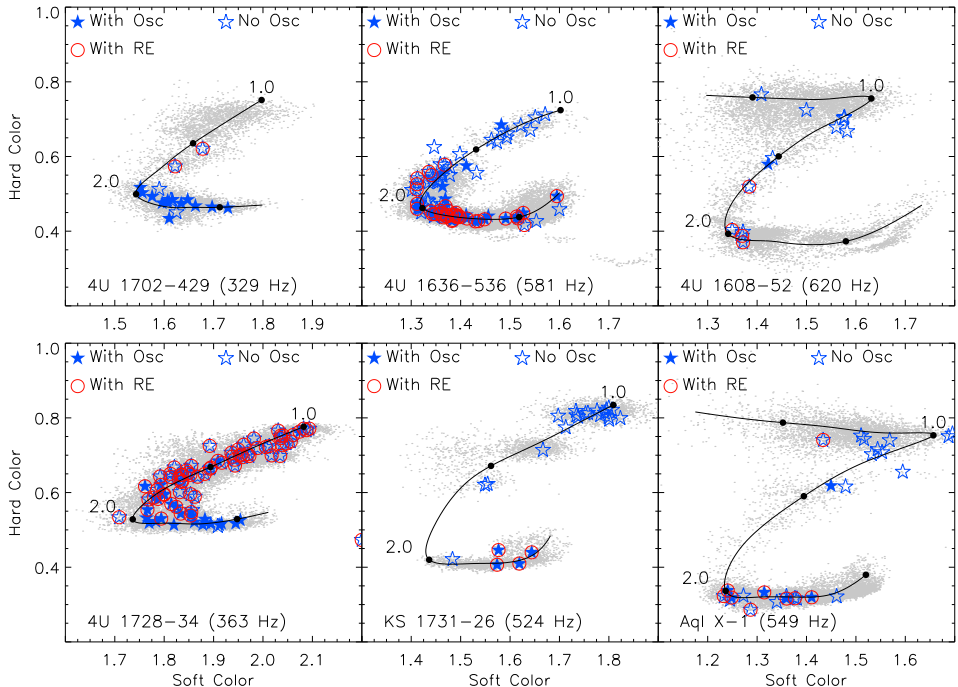,width=0.75\linewidth}}
\caption{Properties of the X-ray bursts as a function of the persistent flux 
for six sources of millisecond burst oscillations. The persistent flux is 
parameterized by soft and hard colors. An increasing accretion rate onto the 
neutron star is
traced as the source moves from the top-left to the bottom-right 
along the Z-track in the color-color diagram. We parameterize the position
along the Z-track 
using the arc length $S_Z$ of the black curve; the arc length is 
normalized to the distance between $S_Z = 1$ at the top-right vertex
and $S_Z = 2$ at the bottom-left vertex. The persistent colors before
the X-ray bursts are indicated with blue stars. Filled stars indicate that
oscillations are detected during the bursts. Bursts with photospheric radius 
expansion are circled in red. Burst oscillations are always 
seen at high accretion rates. However, in the slow sources (leftmost panels)
radius expansion only occurs at low accretion rates, whereas in fast
sources (center and right panels) radius expansion occurs at high 
accretion rates.
}
\label{fig:cc}
\end{figure*}
explained in Galloway et al. (in preparation). As of August 2003, we detected
394 bursts from these 12 sources. 

\subsection{Persistent Accretion Rate onto the Neutron Star}

We first characterized the persistent 
accretion rates onto the neutron stars ($\dot{M}$) using X-ray color-color 
diagrams. Using data with 128 energy channels between 
2--60~keV and 16 s time resolution, we defined soft and hard colors as 
the ratio of the background-subtracted detector counts in the 
(3.6--5.0)/(2.2--3.6) keV and the (8.6--18.0)/(5.0--8.6) keV energy bands, 
respectively. We used 64 s integrations to calculate the colors when the 
source intensity was above 100 counts s$^{-1}$, and 256 s integrations 
otherwise. We corrected the raw colors to account for changes in the 
gain of the PCA using linear terms that were determined by assuming that 
the Crab nebula had a constant count rate in all four bands 
(Muno, Remillard, \& Chakrabarty 2002a).

The resulting color-color diagrams are illustrated using 
the grey points in Figure~\ref{fig:cc}. 
As $\dot{M}$ onto the neutron star increases,
a source moves from the top-left to the bottom-right, roughly tracing
a Z-shaped pattern 
\citep{hv89,mrc02,gd02a}.\footnote{Note that all of the
burst oscillation sources are traditionally classified as so-called 
``atoll'' sources.
They are not to be confused with the so-called ``Z'' sources, 
which are more luminous and do not usually exhibit X-ray bursts 
\citep[see][for a review]{vdk00}.}
The transient systems \aqlxone\ and \novab\ were observed over the largest 
range of accretion rates, and therefore trace a full Z-shaped pattern in 
the color-color diagrams. The sources 
\sevenb, \slowb, \sixb, and \ksxrb\ trace portions 
of the bottom and diagonal branches of the Z-track 
\citep[see also][]{mun00,fra01,vst01}. For these six sources, we 
have parameterized the position on the color-color
diagram using the arc-length ($S_Z$) along the solid curves in 
Figure~\ref{fig:cc} \citep[see][for details of how this is defined]{die00}. 
The remaining sources were observed on only a portion of the Z-track. 
The sources \mxbecl\ \citep{wij02}, \dipb\ \citep{boi00}, SAX~J1748.9$-$2021 \citep{kaa03}, and SAX~J1750.8-2900 \citep{kaa02} were 
observed almost extensively on the bottom portion of the Z-track. 
The persistent pulsars \saxmsp\ and XTE~J1814-338 were only observed
on the top (hard) branch of the Z-track, despite large variations in 
their luminosities. The color-color diagrams for these sources will be 
presented elsewhere (Galloway et al., in preparation).

In order to determine the accretion rate at the time the X-ray burst 
occurred, we calculated the hard and soft colors in the
256~s interval prior to the burst. These are indicated by the blue 
stars in Figure~\ref{fig:cc}. Most of the conclusions that follow
will be based on the six sources in which $\dot{M}$ varies over a wide range.

\subsection{Presence of Millisecond Oscillations}

We searched for millisecond oscillations during the X-ray bursts in data 
recorded with $2^{-13}$ s (122 $\mu$s) time 
\begin{figure*}[thb]
\centerline{\epsfig{file=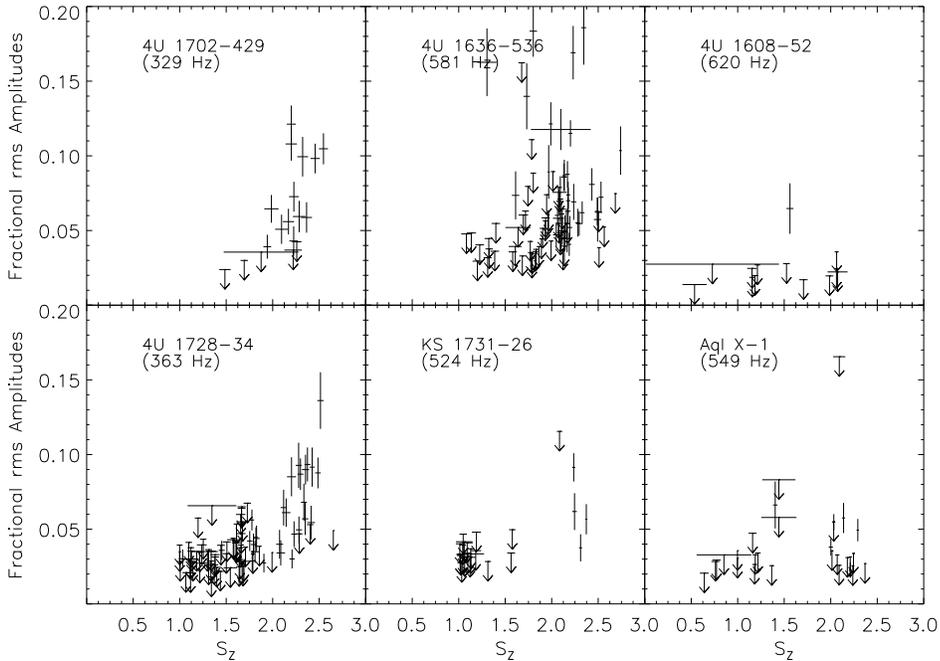,width=0.75\linewidth}}
\caption{
The fractional rms amplitudes of detected millisecond 
burst oscillations, and upper 
limits for non-detections, plotted as a function of position on the 
color-color diagram ($S_Z$). The position on the color-color diagram is 
proportional to the accretion rate onto the neutron star. 
Oscillations are preferentially detected
at higher accretion rates, and likewise the rms amplitudes of the detected
oscillations are systematically larger than the upper limits obtained 
at low accretion rates. This suggests that the millisecond 
oscillations decrease in fractional amplitude as the accretion rate drops.
}
\label{fig:rms}
\end{figure*}
resolution.
We computed fast Fourier transforms of each 1~s interval of data for the 
first 15~s of the burst, and searched for signals within 5~Hz of the 
known spin frequencies. We considered a 
signal to be a detection if it had less than a 1\% probability of occurring 
by chance given the 160 trial frequencies searched for each burst. A 
signal was 
considered significant if it passed any of three tests: (1) having a chance of
$< 6\times10^{-5}$ that it was produced by noise in a single 
trial, (2) persisting for two adjacent time and frequency bins with a chance 
probability of $< (6\times10^{-5})^{1/2}/6 = 1\times10^{-3}$,
or (3) occurring in the first second of a burst with a chance probability 
of $<10^{-3}$. Millisecond oscillations were detected from 137 bursts in 
this manner;
fewer than 5 ($< 3$\%, conservatively assuming the tests were independent 
trials) 
should be spurious detections of noise signals according to 
our selection criteria. Oscillations are detected in all of the bursts
from the millisecond pulsars \saxmsp\ and XTE~J1814$-$338 
\citep{cha03,str03}, but are almost exclusively 
observed at high apparent accretion rates ($S_Z \gtrsim 2$) in the 
remaining LMXBs. We have indicated the X-ray bursts 
with oscillations using filled blue stars in Figure~\ref{fig:cc}. 

We examined why the millisecond oscillations are not detected at low 
accretion rates by comparing the fractional rms 
amplitudes of the detected oscillations to 
the upper limits on the non-detections. The fractional rms amplitude $A$ is 
defined as
\begin{equation}
A = \left( {{P} \over {N}} \right)^{1/2} {{N} \over {N-B}},
\end{equation}
where $P$ is the power in the oscillations, $N$ is the total number of 
counts used to compute the power spectrum, and $B$ is the estimated 
background rate taken from the 16~s prior to the burst \citep{lea83}. 
For the 6 sources in Figure~\ref{fig:cc}, we plot in Figure~\ref{fig:rms}
the amplitudes as a function of position on the Z-track. Each burst is recorded
once in the figure. For detected oscillations, we plot the amplitude of the 
signal with the largest power. For upper limits, we plot the largest 
rms amplitude among the noise signals detected within 5~s after the start of
the burst decay (most detected oscillations are observed at this time;
see Muno \etal\ 2002b). Although the trend is 
not absolute, the upper limits on the non-detected signals at low $\dot{M}$ 
are typically a factor of two smaller than the amplitudes of the 
detected signals at higher $\dot{M}$. This indicates that the fractional 
amplitudes decline as the accretion rates onto the neutron stars decrease.

\subsection{Photospheric Radius Expansion}

We characterized the spectral properties of the X-ray 
bursts by extracting spectra of
the emission at 0.25~s intervals. The data used provided at least 32 energy
channels between 2--60~keV. We accounted for the background and 
non-burst emission using spectra 
from the 16~s prior to each burst \citep[e.g.][]{kul02}. The instrumental
responses for the detector units that were active during each burst
were generated using the FTOOL {\sc pcarsp}. We modeled 
the spectra between 2--25~keV using a blackbody function absorbed by 
the interstellar medium. The interstellar hydrogen column density was 
held fixed at the
mean value determined from a prior fit in which it was allowed to vary.
The model provides an apparent temperature ($T_{\rm app}$) and a
normalization equal to the square of the apparent radius ($R_{\rm
app}$) of the burst emission surface. 

As a proxy for the other properties
of the bursts, we adopt the presence of photospheric radius expansion, 
using the criteria of \citet{gal03}. We considered a source to have 
exhibited radius expansion if (1) $R_{\rm app}$ reached a local maximum 
close to the time of peak flux, (2) $R_{\rm app}$ was observed to be lower
shortly after the flux, with a significance of at least $4\sigma$, and 
(3) coincident with the maximum in $R_{\rm app}$, $T_{\rm app}$ also reached 
a local minimum. X-ray bursts with radius expansion that are identified
with these criteria also tend to have the largest
peak fluxes and fluences, and usually rise and decay most quickly
\citep[][Galloway et al., in preparation]{sb03}. 

The X-ray bursts with radius expansion are indicated with red circles in 
Figure~\ref{fig:cc}. Radius expansion bursts occur preferentially at low  
$\dot{M}$ ($S_Z \lesssim 2$) in the 
sources with $\simeq 300$~Hz spin frequencies in Figure~\ref{fig:cc}
(\sevenb\ and \slowb), but not at high $\dot{M}$ 
($S_Z \gtrsim 2$). 
In contrast, radius expansion occurs predominantly at high $\dot{M}$ 
($S_Z \gtrsim 2$) in the sources with $\simeq 600$~Hz spin frequencies 
in Figure~\ref{fig:cc} (\sixb, \novab, \ksxrb, and \aqlxone). 
The corresponding trends for the persistent millisecond pulsars are unclear, 
because they only trace the top branch of the Z-track on 
the color-color diagram, 
and all the observed bursts exhibit oscillations. 
However, the bursts from the 401~Hz pulsar \saxmsp\ all exhibit radius 
expansion like the fast sources \citep{cha03}, while the bursts from from 
the 314~Hz pulsar 
XTE~J1814--338 all lack radius expansion like the slow sources \citep{str03}.
The trends are less clear for the remaining LMXBs because less
data is available, but they are consistent with those in Figure~\ref{fig:cc}
(Galloway et al. in preparation).

\section{Discussion}

In a previous paper, we found that for neutron stars spinning at 
$\simeq 600$~Hz, millisecond oscillations were observed preferentially 
in thermonuclear X-ray bursts with photospheric radius expansion, 
while for neutron stars spinning at $\simeq 300$~Hz, oscillations 
were equally likely to be observed 
in bursts with and without radius expansion \citep{mun01}. 
The additional data that has entered the public \rxte\ archive in the
past two years demonstrates that this initial distinction can 
be explained as a combination of two effects. First,
in all of the sources 
oscillations are preferentially detected when the accretion rates onto
the neutron stars are high. 
Second, the X-ray bursts that occur at high accretion 
rates in the fast rotators exhibit radius expansion, but do not in the 
slow rotators.

\subsection{Millisecond Oscillations as a Function of $\dot{M}$} 

The detectability of millisecond burst oscillations is clearly not determined 
by the properties of the X-ray bursts that they appear in, because if 
one considers the 
entire sample of sources in Figure~\ref{fig:cc}, bursts with and without 
photospheric 
radius expansion are equally likely to exhibit oscillations. The properties of 
the X-ray bursts are probably the best indicators of the conditions in the 
burning layer, while the low amplitudes of the oscillations 
\citep[5--10\% rms;][]{moc02} suggest that they are only a minor side-effect 
of the burning. Therefore, since the the properties of the bursts at low 
accretion rates are so dramatically different in the fast and slow rotators, 
it seems unlikely that there is some subtle change in the mechanism producing
the oscillations that prevents them from being observed at low $\dot{M}$
from all of the sources.

Therefore, we suggest that the mechanism producing the millisecond
oscillations always 
operates during X-ray bursts, but that their amplitudes are attenuated at low 
$\dot{M}$ by a mechanism external to the burning layer. 
The amount of attenuation required is quite small, since the fractional
amplitudes of the detected signals are on average only a factor of 2 larger 
than the upper limits on the non-detections (Figure~\ref{fig:rms}).
The most likely source of attenuation is a corona of electrons that scatters
photons from the surface of the neutron star \citep[e.g.][]{bl87,mil00}. 
A corona of optical
depth $\tau \approx 3$ would be sufficient to attenuate the burst oscillations
by a factor of 2 \citep[e.g.,][]{mil00}. Such 
a corona of electrons is thought to produce the high-energy 
power-law tail that is often present in the X-ray spectra of LMXBs, by 
inverse-Compton scattering thermal photons form the neutron star 
\citep[e.g.,][]{bar00,gd02b}. This
power-law tail is only observed at low $\dot{M}$, and is inferred to originate 
from a corona
with temperature $kT \ga 20$ keV and optical
depth $\tau approx 3$ \citep{bo02, gd02b,mc03}. Therefore, this corona could 
be the reason that millisecond burst oscillations are 
not observed at $S_Z \lesssim 2$ in Figure~\ref{fig:cc}.\footnote{We note,
however, that at high accretion rates the same authors infer the presence
of a cooler ($kT \la 5$ keV), more optically thick ($\tau \ga 5$) Comptonizing 
corona. Our interpretation would require that at high $\dot{M}$ this corona is 
geometrically arranged so as not to intercept photons from the surface of 
the neutron star.}

The two persistent
millisecond pulsars provide the only exceptions to the above trend:
all of the burst oscillations, and
all of the X-ray bursts, are observed in the hard portion of the Z-track on
the color-color diagram that corresponds to low $\dot{M}$. The main 
proposed difference
between the millisecond pulsars and the other bursters is that the 
former have stronger magnetic fields \citep{cha03}. It is 
therefore plausible that oscillations are observed at low inferred $\dot{M}$
for the pulsars because magnetic effects either enhance the
oscillation amplitudes, or supress a scattering corona.

\subsection{X-ray Burst Properties as a Function of $\dot{M}$}

The properties of X-ray bursts also are correlated with the  
accretion rate onto the neutron star, but in a manner that additionally 
depends on the neutron star rotation rate.
We expect the burst properties to be determined by two 
factors \citep[Fujimoto, Hanawa, \& Miyaji 1981;][]{fl87,bil00}. 
First, as $\dot{M}$ increases, the 
temperature at the burning layer also increases, and thus the column density
of helium required to trigger a burst generally decreases. Therefore, if the 
accretion is spherically symmetric, X-ray bursts that occur 
at high $\dot{M}$ should be weaker.

Second, there is a competition between 
how quickly a sufficient column density is accumulated such that 
helium burning is 
unstable, and how quickly hydrogen in the accreted material can be 
stably fused into helium. The local accretion rate per unit area 
($\dot{m}$) actually
drives the competition, but the accretion is generally assumed to occur
with spherical symmetry.
At values of $\dot{m}$ thought to correspond to the lower end of those 
commonly observed from bursting LMXBs
(equivalent to global accretion rates of 
$0.01\dot{M}_{\rm Edd} < \dot{M} < 0.05\dot{M}_{\rm Edd}$), H
is burned into He faster than it can be accreted, so the X-ray burst occurs 
from pure He fuel. As $\dot{m}$ increases 
($0.05\dot{M}_{\rm Edd} < \dot{M} < \dot{M}_{\rm Edd}$), H is 
accreted faster than it can
be burned into He, so the bursts occur from mixed H/He
fuel.\footnote{We note that at $\dot{M} < 0.01\dot{M}_{\rm Edd}$ hydrogen 
burning is unstable, which should also produced mixed H/He bursts. However,
in this regime the 
expected recurrence times (30~h to 10 days; see Bildsten 2000; Narayan \& 
Heyl 2003) are significantly longer than those observed 
(2--10~h; see Cornelisse \etal\ 2003). This regime is unlikely to apply to
the current data.}
The X-ray burst properties change between these regimes because the relative 
amount of H and He in the fuel determines how rapidly
the nuclear energy is released during the bursts. 
Helium burns via a strong 
triple-$\alpha$ process that releases energy quickly, so the low-$\dot{m}$
He bursts are more likely to exhibit radius expansion. In contrast, 
H serves to moderate the He burning at the start of the burst, and only
burns through a slow $rp$-capture process onto the products of 
He burning at the end of a burst. Therefore, the high-$\dot{m}$ 
mixed H/He bursts
should last longer, and be less likely to exhibit radius 
expansion.

As a result of these two effects, as the global accretion rate onto
the neutron star increases, the 
X-ray bursts should become weaker and less likely to exhibit radius expansion.
This is the case for the slow rotators, but in the fast rotators the 
bursts are {\it more likely} 
to exhibit radius expansion at high $\dot{M}$.
One possible explanation for this is that the accretion is not spherically
symmetric \citep{bil00}. In particular, if the local accretion rate ($\dot{m}$)
{\it decreases} as the global rate ($\dot{M}$) increases in the fast 
rotators, then the high-$\dot{m}$ bursts with radius 
expansion could occur at low $\dot{M}$, and
the low-$\dot{m}$ bursts without
radius expansion could occur at high $\dot{M}$. 
In contrast, the change in burst properties in the slow rotators appears
consistent with a local $\dot{m}$ that increases as the global $\dot{M}$
does. It is then possible that the rotation rate of the neutron star 
influences how accreted material spreads over its surface, either through
a lower effective surface gravity or a stronger Coriolis force in the 
fast rotators.; \citep[e.g.][]{bil00,slu02}. 

Although we have determined observationally that the rotation rate
of a neutron star influences how the properties of thermonuclear 
X-ray burst change with the 
accretion rate onto the neutron star, it is not clear what causes the 
observed correlations. Further progress should be made by studying this 
sample of sources to see how the burst time scales, 
peak fluxes, fluences, and recurrence times change with the accretion rates.

\acknowledgments{We thank R. Wijnands for providing the flux history of 
\saxmsp, and J. Hartman for reducing the data on XTE 1814--338. We also 
thank the referee for helpful questions and suggestions.
This work was supported by NASA under contract 
NAS 5-30612 and grant NAG 5-9184, and through a Hubble Fellowship
to MPM from the Space Telescope Science Institute, which is operated
by the Association of Universities for Research in Astronomy, Inc.,
under NASA contract NAS 5-26555.}

\end{document}